\documentclass[prl,aps,11pt]{revtex4-1}
\usepackage{graphicx}
\usepackage{amsmath}
\usepackage{graphicx}

\begin{document}
\title{Accounting for backflow in hydrodynamic-simulation interfaces}
\author{Scott Pratt}
\affiliation{Department of Physics and Astronomy \\and National Superconducting Cyclotron Laboratory\\
Michigan State University\\East Lansing, MI 48824~~USA}
\date{\today}
\begin{abstract}
Methods for building a consistent interface between hydrodynamic and simulation modules is presented. These methods account for the backflow across the hydrodynamic/simulation hyper-surface. The algorithms are efficient, relatively straight-forward to implement, and account for conservation laws across the hyper-surface. The methods also account for the spurious interactions between particles in the backflow and other particles by following the subsequent impact of such particles. Since the number of altered trajectories grows exponentially in time, a cutoff is built into the procedure so that the effects of the backflow are ignored beyond a certain number of collisions. 
\end{abstract}

\maketitle
\section{Introduction and Motivation}

Relativistic viscous hydrodynamics is a popular choice for modeling high-energy heavy-ion collisions. Hydrodynamics is appropriate when collisions are sufficiently rapid to keep the various species moving with a single collective velocity and local kinetic temperature, and to keep the stress-energy tensor sufficiently isotropic to warrant a viscous treatment. However, these conditions are lost near the end of the reaction, when the various hadrons begin to cool separately and move with different collective velocities \cite{Sorge:1995pw,Pratt:1998gt}. The final stage of the reaction and the decoupling are then best modeled with a microscopic simulation, which in the limit of many particles, or with a high over-sampling, becomes equivalent to a Boltzmann description. Often these simulations are referred to as hadronic cascades.

Particles are emitted into the hadronic cascade through a hyper-surface separating the hydrodynamic region from the cascade. If the phase space density of a particle's of momentum ${\bf p}$ at the boundary is $f({\bf p})$, the number of particles emitted into the simulation side of the boundary through a hyper-surface element is given by the Cooper-Frye formula \cite{Cooper:1974mv},
\begin{equation}
\label{eq:cooperfrye}
dN=\frac{d^3p}{(2\pi)^3}\frac{p\cdot \Omega}{E_p}f({\bf p}).
\end{equation}
Here, $\Omega_\mu$ is a small hyper-surface element. Variations of the Cooper-Frye formula have been applied to numerous hybrid models \cite{Nonaka:2006yn,Gale:2012rq,Song:2011hk,Novak:2013bqa}. In each of these approaches the Boltzmann equation is solved by sampling techniques, i.e., rather than storing information for each phase-space element $d^3pd^3r/(2\pi)^3$, one follows the evolution of sample particles chosen consistently with the phase space density. For sampling ratios of one, the cascade is a one-to-one simulation of the hadronic stage. For higher sampling ratios, the models approach the limit of a Boltzmann equation. Since hadronic cascades are modeling the low-density stage at the end of the collision where velocity gradients are reduced, sampling ratios of unity are nearly indistinguishable from the Boltzmann limit. This is in contrast to the case of simulating the early partonic stage, where sampling ratios need to be of order 10 or more to approach the Boltzmann limit \cite{Molnar:2001ux,Cheng:2001dz}. Whether the sampling ratios are unity or not, one needs to generate particles into the cascade code consistently with the hydrodynamic description at the hyper-surface.

For a time-like element, $\Omega^2>0$, one can consider the emission from a frame where the emission is simultaneous across the element and $d\Omega_0$ represents a volume element undergoing sudden emission. For a space-like element, one can choose a frame where the surface is stationary. In this frame $\Omega_i$ represents the area of the element multiplied by the time of the emission. Depending on the hyper-surface element, $\Omega_\mu$, and the particle's momentum $p^\mu$, the number $dN$ can be either positive or negative. The positive contribution describes particles being fed into the cascade, whereas the negative contribution represents the backflow, i.e. those particles which leave the cascade region, cross the interface, and enter the hydrodynamic domain. Both the positive and negative contributions are necessary if energy, momentum and charge are to be conserved across the interface. Similar issues to those being discussed here in the context of a hydrodynamics/cascade interface have also been studied with regards to coupling a hydrodynamics directly to the vacuum, instantaneous freezeout, see for example \cite{Anderlik:1998et}. When coupling directly to the vacuum the stress-energy tensor is discontinuous, while with coupling to a Boltzmann code, one can aim for a continuous description if the interface is performed at a sufficiently high density that collision rates justify the hydrodynamic description.

Most hadronic cascades ignore backflow. Such codes apply a step function, $\Theta(p\cdot\Omega)$, to Eq. (\ref{eq:cooperfrye}), and don't erase particles from the simulation that flow backward across the boundary. For modeling of particles at mid-rapidity, it has been shown that the backflow is on the order of one half of one percent \cite{Huovinen:2012is,Pratt:2008sz}. However, it has been reported that the error grows to the level of several percent away from mid-rapidity \cite{Huovinen:2012is}. The purpose of this short paper is to introduce a  method for correcting for backflow. The approach accounts for the backflow through the production and propagation of negative-weight particles. Local conservation laws are exactly conserved in the limit of high sampling. The theoretical underpinnings of this approach are described in the next two sections. A hadronic cascade was altered to incorporate these changes, and a brief evaluation of the behavior is presented in the subsequent section along with conclusions about the approach. The appendix provides a description of the algorithm used to numerically sample the particle flow across the hydrodynamic/cascade interface.

\section{Cooper-Frye Formalism}

In order to apply the Cooper-Frye formalism of Eq. (\ref{eq:cooperfrye}) one needs to first generate a list of hyper-surface elements, $\Omega_\mu^I$, where the index $I$ denotes each small element. Formally, $\Omega_\mu^I$ can be described as the portion of the hyper-surface that falls within a small four-volume element, $\int_Id^4x$. The surface is described by a criteria based on local properties, such as the temperature or density, and can be written parametrically as,
\begin{equation}
C(x)=C(T(x),\rho(x))=0,
\end{equation}
where the function $\rho$ could be any quantity propagated through the hydrodynamic evolution, such as the particle or energy density. For instance, if the interface is chosen to occur at fixed temperature, $T_b$, the location of the surface would be described by $C=T(x)-T_b=0$. The function should be defined such that one is inside the hydrodynamic region if $C>0$ and outside if $C<0$. The hyper-surface element is then
\begin{equation}
\Omega_\mu^I=-\int_Id^4x~\partial_\mu \Theta(C(x))=-\int_Id^4x~\delta(C(x))\partial_\mu C(x),
\end{equation}
where $\Theta$ is a step function. For example, if $C$ is defined by a surface of constant temperature, $\Omega_\mu^I$ is parallel to $-\partial_\mu T$.

For time-like elements, $\Omega^2>0$, one can always consider the element in a frame where $\Omega$ points purely in the time direction. In this frame the emission is simultaneous across the hyper-surface, and $\Omega_0$ represents a small volume element. If $\Omega_0>0$ the criteria $C$ is falling with time, e.g. the temperature is falling, and particles are being emitted into cascade region. The majority of the emission in heavy-ion collisions comes from such elements. For $\Omega_0<0$, the criteria is rising, and the hydrodynamic region is absorbing the volume element. This can be understood by considering the sign of the $p\cdot \Omega$ term in the Cooper-Frye formula. Having $\Omega_0<0$ is unusual in a time-like element, because it is difficult to find an area where the density or temperatures are rising. Even with lumpy initial conditions, the longitudinal expansion is driving the densities downward even in regions where two lumps are expanding into one another. 

For space-like elements, $\Omega^2<0$, one can always find a boost and rotation which point $\Omega_\mu$ along the positive $x$ axis. In this frame the emission surface is temporarily stationary, and the emission $dN$ is positive for particles with $p_x>0$ and negative for $p_x<0$. If the flow velocity in this frame, $u_x$, is greater than zero, there will more positive contribution than negative contribution. For explosive collisions this is usually the case. The negative contributions represent those particles flowing into the hydrodynamic region, i.e., the backflow.  

Both the positive and negative contributions are required to conserve charge. For instance, the charge that travels through an element, from the hydrodynamic to the cascade region between times $t_1$ and $t_2$, is
\begin{eqnarray}
\Delta Q&=&\sum_I \Omega_\mu^I j^\mu\\
\nonumber
&=&\int_{t_1<x_0<t_2} d^4x~j^\mu\partial_\mu \Theta(C(x))\\
&=&-\int_{t_1<x_0<t_2} d^4x~\Theta(C(x))\partial\cdot j
+\int_{x_0=t_1} d^3x~j_0(x)\Theta(C(x))-\int_{x_0=t_2}d^3x~j_0(x)\Theta(C(x)).
\end{eqnarray}
The first term vanishes from current conservation and the last two terms describe the difference between the net charges in the $C(x)>0$ region at the two times. The net current density for all the hadron species $a$ can be written in terms of the phase space density as
\begin{equation}
j^\mu(x)=\sum_a Q_a\int \frac{d^3p}{(2\pi)^3}~\frac{p^\mu}{E_p}f_a(p,x).
\end{equation}
Since the contributions from all momenta are required to construct the conserved current density at an position $x$, one cannot throw away the contributions to the Cooper-Frye formula from those momenta with $p\cdot\Omega<0$ without violating current conservation. Similar expressions can be derived for the energy and momentum, thus showing that the negative contributions to the Cooper-Frye expressions are essential if one wishes to satisfy any of the conservation laws.

There exist numerous algorithms for finding the hyper-surface, depending on whether the initial conditions are lumpy or smooth, whether the calculation is one-dimensional, two-dimensional, or three dimensional, and depending on what sort of mesh is used to model the hydrodynamic expansion. A particularly robust algorithm that works for three-dimensional systems of arbitrary topology is described in \cite{Huovinen:2012is}.

\section{Methods for Accounting for backflow}

The output of hydrodynamic codes is a list of the hyper-surface elements, $\Omega^I$, their space-time coordinates, and any additional information required to reconstruct the phase space density, e.g., the collective flow, temperature, densities, and anisotropies of the stress-energy tensor. For each $d^3p$ within each $d\Omega^I$, one can calculate the probability of creating a particle, $dN$. One then decides to create a particle with probability $dN$. An efficient algorithm for doing this is presented in the appendix. The issue of backflow occurs when $dN$ is negative. In the field of relativistic heavy ions, two approaches have been applied thus far. The first approach is simply to neglect the negative contribution. As mentioned in the introduction, this violates energy, momentum and charge conservation at the one percent level at mid-rapidity, and higher away from mid-rapidity.

A second method for handling the backflow has been to erase those particles from the cascade simulation that cross back into the hydrodynamic region. This requires storing the location of the hyper-surface from the hydrodynamic module for consideration by the cascade. This method would satisfy the conservation laws as long as the phase space density is continuous across the boundary. A strong discontinuity would suggest that the hydrodynamic treatment was not justified for the densities or temperatures that determined the interface, or that the viscous corrections to the stress-energy tensor in the hydrodynamic treatment is not consistent with the dynamics of a hadron gas. Choosing a higher interface density or temperature, or better accounting for viscous effects, might solve the problem. For the purpose of this paper, we will assume that the phase space density is continuous, though with the recommendation that the continuity be better studied in the future. A more daunting problem with this approach comes from the book-keeping required to decide whether particles crossed the boundary. Hydrodynamic approaches tend to use very small cells, on the order of 0.1 fm. Particles would typically cross hundreds of such cell boundaries, with a very small fraction of such cell boundaries representing an interface with the hydrodynamic treatment. Although there are a variety of algorithms used in cascades, it would seem that the majority of the numerical cost in performing the cascades might then be applied to tracking the backflow. Given the field's interest in fluctuating (lumpy) initial conditions, the topology of the breakup surface could be complex, and could vary event-to-event. This makes it difficult to find a robust algorithm that avoids constantly checking for interface crossings. This approach was applied in \cite{Pratt:2008sz} but that was for a smooth, azimuthally symmetric, boost-invariant hydrodynamic treatment, where the hyper-surface could be represented by giving the radius of the transition hyper-surface as a function of time.

A third scheme is presented here. Instead of tracking particles relative to the interface, this scheme simulates the evolution of particles emitted with the negative weights, $p\cdot\Omega$, from Eq. (\ref{eq:cooperfrye}). A weight of negative unity is originally assigned to such particles, whereas a weight of positive unity is assigned to particles emitted with $p\cdot\Omega>0$. Since the negative-weight particles are a small fraction of the overall particles, the additional numerical cost should be on the order of a few percent. The particles are then evolved through the cascade, but with the products of each collision being assigned weights as described below. When analyzing the final products for their impact on specific observables, each particle would contribute proportional to its weight. For instance, when incrementing a bin used to calculate spectra, a particle with weight of -1 would reduce the bin count by one.

If the phase space density is continuous across the interface, the particles being scattered from the cascade region back into the hydrodynamic region should exactly cancel the negatively weighted particles emitted according to the Cooper-Frye formula. Even if the phase space density turns out to be discontinuous, this method will still satisfy all the conservation laws.

It is propagating the weights through the subsequent collisions where the weighting becomes complicated. If a pair of incoming particles each has a weight of positive unity, both incoming particles are removed from the particle list after the collision (a weight of zero is assigned to the incoming particles) and a weight of positive unity is assigned to each of the collision products. However, if the $N$ incoming particles have arbitrary weights, $w_{i,1}\cdots w_{i,N}$, both the new outgoing particles and incoming particles may need to be assigned post-collision weights if the conservation laws are to be satisfied. If the new weights assigned to the incoming particles are non-zero, those particles cannot be deleted from the particle list after the collision.

The weights for the $M$ outgoing particles (those that are newly created or are scattered into new momentum states), are labeled $w_{f,b}$, and are all set to the product of the weights of the incoming particles,
\begin{eqnarray}
\label{eq:weights1}
w_{f,b}&=&W,~~b=1,M,\\
W&\equiv&\prod_{a=1}^N w_{i,a}.
\end{eqnarray}
The weights for the scattered or newly created particles are all equal to the product of the incoming weights. This ensures that if any of the incoming weights are changed by a factor $F$, that the weight of each of the outgoing products is changed by the same factor. Instead of being erased, the incoming particles are assigned post-collision weights $w'_{f,a}$,
\begin{equation}
\label{eq:weights2}
w'_{f,a}=w_{i,a}-W, ~~a=1,N.
\end{equation}
If all the incoming weights are unity, $w'_{f,a}=0$ for all $a$, and the incoming particles can be erased. Assigning non-zero post-collision weights to the incoming particles as prescribed in Eq. (\ref{eq:weights2}) ensures that conservation laws are enforced in each collision,
\begin{equation}
\sum_{a=1,N} w_{i,a}Q_a=\sum_{b=1,M}w_{f,b}Q_b+\sum_{a=1,N}w'_{f,a}Q_a,
\end{equation}
given that the charge is conserved in the reaction,
\begin{equation}
\sum_{a=1,N}Q_a=\sum_{b=1,M}Q_b.
\end{equation}
The conserved charge $Q$ could just as easily refer to a continuous variable like momentum or energy.

For the several simple process involving $2\rightarrow M$ scatterings or $1\rightarrow M$ decays, the weights of the outgoing particles are shown in Eq.s \ref{eq:weights}. Each incoming particle is labeled $(p,w)$, where $p$ refers to all information about the particle and its trajectory, while $w$ is its weight. The outgoing particles are similarly referenced, with $q$ referring to the particle's information. If all weights were unity, the weights could be ignored and the reactions would be described as $p_1,p_2\rightarrow q_1\cdots q_M$. One can understand the weightings for each line of Eq. (\ref{eq:weights}) from Eq. (\ref{eq:weights1}). For example, one can consider a case where a positive-weight particle collides with a negative-weight particle, $w_{i,1}=1, w_{i,2}=-1$. The positive-weight particle should not have scattered since the negative-weight particle should not be there, plus the negative-weight particle is in principle canceling the effect of a spurious positive-weight particle that should have been erased from cascade for having re-entered the hydrodynamic region. Thus, not only should particle-1 not scatter, one needs to correct for the fact that such particles sometimes scatter spuriously, which means the forward going track weight needs to go from 1 to 2. The scattered tracks are given negative weight so that the cancel the effect of particle-1 scattering off a spurious track. Similar arguments follow for each row in the list. If the phase space density is continuous across the border, the original negative weight tracks should exactly cancel the positive-weight tracks that would have been erased when reentering the hydrodynamic region if one were applying the second method described earlier.
\begin{eqnarray}
\label{eq:weights}
W&\equiv&\prod_{a=1}^N w_a\\
\nonumber
(p_1,w_1),\cdots,(p_N,w_N)&\rightarrow& (p_1,w_1-W),\cdots,(p_N,w_N-W),~~(q_1,W),\cdots,(q_M,W)\\
\nonumber
(p_1,+1)&\rightarrow&(q_1,+1),\cdots(q_M,+1)\\
\nonumber
(p_1,-1)&\rightarrow&(q_1,-1),\cdots(q_M,-1)\\
\nonumber
(p_1,+1),(p_2,+1)&\rightarrow& (q_1,+1),\cdots(q_M,+1)\\
\nonumber
(p_1,+1),(p_2,-1)&\rightarrow&(p_1,+2),~~(q_1,-1),\cdots(q_M,-1)\\
\nonumber
(p_1,-1),(p_2,-1)&\rightarrow&(p_1,-2),(p_2,-2),~~(q_1,+1),\cdots(q_M,+1)
\end{eqnarray}

Implementing the procedure described in Eq. (\ref{eq:weights}) turns out to be non-tenable due to the growth in the weights of the outgoing products. For example, if two particles with weight 2 collide, the scattered particles would have weights of 4. In practice, this lead to exponentially growing weights as a function of the number of collisions. In addition to the growing weights, the number of trajectories being considered increases since the incoming particles are often reweighted, rather than deleted. For a central Au+Au collision at the highest RHIC energy, final-state weights would often exceed the numerical range of the computer, and the number of tracks would exceed the available memory. Even if the simulation were allowed to finish, the high-weighted tracks would overwhelm the answer with noise. Thus, the procedure needs to be modified so that the effects of scattering with non-unity-weighted particles is regulated.

To limit the growth and associated noise of heavily weighted tracks, the procedure described above was modified. Particles were first divided into two sets: base particles and backflow tracer particles. The base particles are created and evolve exactly as if the backflow particles had never been included in the cascade. The tracer particles begin as the set of negative-weighted particles representing the backflow. They also include all the trajectories required to trace the influence of the backflow. Tracer particles are allowed to interact with base particles, but are not allowed to collide with one another. For this reason the effect of the backflow is handled correctly to linear order in the backflow. Since the fraction of backflow particles is unlikely to exceed a few percent, the error associated with this approximation should be on the order of a tenth of a percent or less. Charge, energy and momentum is conserved in each collision as well as in the generation of the particles through the hydrodynamic/cascade boundary.

All of the base particles are assigned weights of positive unity, whereas the tracer particles are allowed to have weights of $\pm 1$. The weights for scattering processes needs to be assigned consistently with those given in Eq. \ref{eq:weights}. However, because collisions between two tracer particles is neglected, the list of interactions is abbreviated and given in Eq. (\ref{eq:altreactionlist}). For a collision between incoming particles whose momenta, position and charges are labeled by $p_1$ and $p_2$ while the outgoing particles are denoted by $q_1,\cdots q_M$, the cascade must consider three cases. In the first case both particles are base particles and have weights of positive unity. Such particles are described by the notation $(p_i,+1,{\rm base})$. In the second case, one of the particles is a tracer particle with positive weight, $(p_i,+1,{\rm tracer})$, and in the third case one has a tracer particle of negative weight.
\begin{eqnarray}
\label{eq:altreactionlist}
(p_1,+1,{\rm base}),(p_2,+1,{\rm base})&\rightarrow&(q_1,+1,{\rm base}),\cdots (q_m,+1,{\rm base}),\\
\nonumber
(p_1,+1,{\rm base}),(p_2,+1,{\rm tracer})&\rightarrow&(p_1,+1,{\rm base}),(p_1,-1,{\rm tracer}),(q_1,+1,{\rm tracer}),\cdots (q_m,+1,{\rm tracer}),\\
\nonumber
(p_1,+1,{\rm base}),(p_2,-1,{\rm tracer})&\rightarrow&(p_1,+1,{\rm base}),(p_1,+1,{\rm tracer}),(q_1,-1,{\rm tracer}),\cdots (q_m,-1,{\rm tracer}).
\end{eqnarray}
For decays, $p\rightarrow q_1,\cdots,q_M$, the outgoing decay products all have the same weights as the incoming particle, and are all either base or all tracer particles depending on whether $p$ is a base or tracer particle. For decays, the original track is deleted.

From inspecting the three reactions in Eq. (\ref{eq:altreactionlist}) one can see that the evolution of the base particles is exactly the same as it would be without the tracer particles, as there is no change to the base particles occurring when the base particle collides with a tracer particles. The tracer particles for the final states are all added into the equations on the right-hand sides of the last two expressions in Eq. (\ref{eq:altreactionlist}) to make the reactions consistent with the weights in Eq. \ref{eq:weights}. For the two reactions involving tracer particles there are two outgoing particles that differ only by their weights or by whether they are tracer particles. Since having two particles with the same trajectory would cause numerical problems, the second particle is translated randomly in relative rapidity in the case of Bjorken boost invariance, or if boost invariance is not implemented, moved by a small random step in coordinate space.

This procedure correctly reproduces the evolution described by the algorithm described in Eq. \ref{eq:weights} to first order in the backflow. However, for the $2\rightarrow 2$ processes described in Eq. (\ref{eq:altreactionlist}) that involve an incoming tracer particle, the number of tracer particles triples from one to three. Thus, if a large number of collisions occur the number or tracer particles quickly grows and the method can become noisy. For that reason, the scattering of tracer particles is curtailed once a tracer particle and its ancestors have suffered a number of collisions $N_{\rm max}$. This is accomplished by storing a number $n_{\rm coll}$ for each tracer particle. Then, when a incoming tracer particle scatters according to the list in Eq. (\ref{eq:altreactionlist}), the number is incremented by one and assigned to all the tracer products of the final state.

\begin{figure}
\centerline{\includegraphics[width=0.6\textwidth]{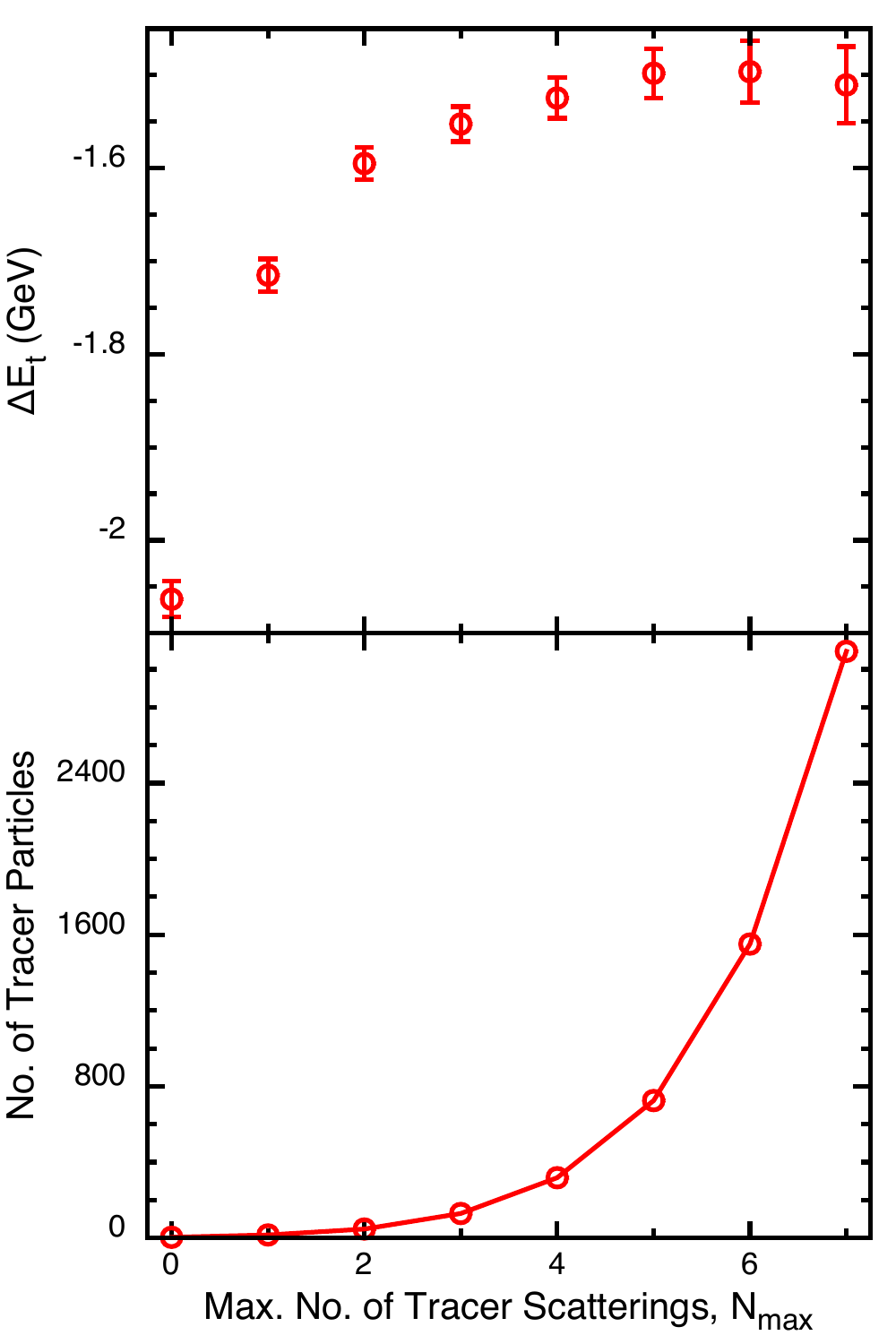}}
\caption{\label{fig:fake}
From a simulation of the central two units of rapidity in a 100$A$ GeV on 100$A$ GeV Au+Au collision, the number of tracer particles, the additional particles required to model the effects of the backflow, is shown in the lower panel as a function of $N_{\rm max}$, which limits the amount of scattering the tracer particles can undergo. For $N_{\rm max}=0$ backflow particles never scatter and only 4.5 particles are required. However, the effect of a scattering spreads over more and more particles as $N_{\rm max}$ is increased, and by the time $N_{\rm max}=7$, the number of tracer particles is greater than the number of actual particles. The lower panel shows the net transverse energy carried by tracer particles in the final state. The effect saturates for large $N_{\rm max}$. However, large $N_{\rm max}$ is numerically more expensive and can introduce more statistical noise into the analysis.
}
\end{figure}

The sensitivity to the cutoff $N_{\rm max}$ is shown in Figure \ref{fig:fake}. The figure shows results from a simulation of 100$A$ GeV on 100$A$ GeV Au+Au central Au+Au collisions at RHIC. The hadronic cascade, B3D \cite{Novak:2013bqa}, modeled a longitudinally boost invariant system by describing hadrons with spatial rapidities between -1 and 1 with cyclic boundary conditions. The algorithm for modeling backflow described above was incorporated into the model with several values of $N_{\rm max}$. For $N_{\rm max}$ set to zero, the backflow particles were created, but were not allowed to collide. After decays, the average number of backflow particles per collision was approximately 4.5 per event, a small fraction of the over 2300 final-state base particles. A typical hadron might collide two to three times per cascade event but backflow particles were likely to collide significantly more often due their being emitted into the dense hydrodynamic region. Since the number of tracer particles triples with each collision, the number of tracer particles rises rapidly as a function of $N_{\rm max}$, as shown in the lower panel of Fig. \ref{fig:fake}, and by the time $N_{\rm max}=7$ the number of final-state tracer particles exceeds the number of final-state base particles.

For $N_{\rm max}=0$, the $\approx 4.5$ backflow particles carried nearly 2.2 GeV of transverse energy, which subtracts from the net transverse energy of the event due to their negative weight. This is denoted by $\Delta E_t$,
\begin{equation}
\Delta E_t\equiv\sum_{{\rm tracers}~i} w_i\sqrt{m_i^2+p_{i,t}^2},
\end{equation}
and is shown in the upper panel of Fig. \ref{fig:fake} as a function of $N_{\rm max}$. Once $N_{\rm max}>0$, tracer particles can have both positives and negative weights. The net transverse energy carried by the tracer particles then involves a large cancellation and can become noisy for large $N_{\rm max}$ due to the large number of tracer particles. By letting the tracer particles scatter, the net transverse energy carried by the tracer particles is reduced by the longitudinal work done in the expansion. The quantity $\Delta E_t$ saturates for larger $N_{\rm max}$. The bulk of the correction for scattering is accounted for with just a few scatterings. For example, by setting $N_{\rm max}=3$, one simulates nearly 90\% of the change of $\Delta E_t$ with little additional computational cost since the average number of tracer particles is only on the order of 100.

\section{Summary}

The backflow contribution from a Cooper-Frye interface between hydrodynamic and Boltzmann/Cascade models can indeed can be accounted for with the methods above. The method was shown to correctly account for the backflow in the limit that the backflow is sufficiently small so that the effects are linear in the backflow. Additionally, a second approximation can be applied that limits the simulation of backflow to a finite number of scatterings. Most of the secondary scattering effects occur in the first few scatterings, which allows one to impose the cutoff for small $N_{\rm max}$. This prevents the method from consuming significant additional numerical resources or from adding significant statistical noise to the output. Even with $N_{\rm max}=0$, charge, energy and momentum flow through the interface is satisfied.

This work was mainly motivated by the desire to find a method to account for backflow without requiring disproportionate resources to describe an effect whose impact is modest at best. At mid-rapidity in high-energy heavy-ion colliisons, the number of particles traveling back into the hydrodynamic region is only a fraction of a percent of the total number of particles. For slightly lower energies, or away from mid-rapidity, that fraction might increase, though is unlikely to exceed a few percent. The method described here seems ideal for such cases.

The method was implemented into the Cooper-Frye generator and cascade used in the code B3D. This code was designed to perform the same tasks as the hadronic cascade URQMD, but to be better suited for a high-energy and to be faster. B3D can generate particles from the hypersurface using the algorithms described in the appendix, and perform the hadronic cascade in approximately 0.25 seconds on a single core in a typical laptop. Adding backflow corrections with $N_{\rm max}=3$ slowed the code down by 3\% and increased the size of the final-state data files by 5\%. The results of Fig. \ref{fig:fake} show that using $N_{\rm max}=3$ allows one to calculate the effects of the backflow to the $\sim 10\%$ level. Given that backflow is a small effect, this may be more than sufficient for most purposes.

\section{Appendix: Algorithm for Generating Particles}

For a small hyper-surface element $\Omega_\mu$, the number of particles emitted is
\begin{eqnarray}
\label{eq:weightdef}
dN&=&\frac{p\cdot\Omega}{(2\pi)^3}\frac{d^3p}{E_p}f({\bf p})\Theta(p\cdot\Omega)\\
\nonumber
&=&w(p)\Omega_{\rm opt}\frac{d^3p'}{(2\pi)^3}f'({\bf p}'),\\
\nonumber
w(p)&=&\frac{\Omega\cdot p}{\Omega_{\rm opt}(u\cdot p)}\Theta(p\cdot\Omega),\\
\nonumber
\Omega_{\rm opt}&=&\Omega\cdot u+\sqrt{(\Omega\cdot u)^2-\Omega^2}.
\end{eqnarray}
Aside from the factor of $w(p)$, the second line looks like the usual thermal emission. The primed quantities refer to momenta and phase space densities measured in the reference frame of the fluid, i.e. the frame where $u\cdot p=E'$. This formula works for any $\Omega_{\rm opt}$, but for a Monte Carlo procedure it must be chosen large enough that $w(p)$ never exceeds unity for any $p$. Efficiency is lost if it becomes larger than necessary. The choice above is the optimum value as it corresponds to the smallest acceptable value that keeps $w(p)\le 1$. 

The strategy is to\vspace*{-6pt}
\begin{enumerate}\itemsep=0pt
\item Choose particles according to a static thermal distribution with volume $\Omega_{\rm opt}$. The quantity $f$ should incorporate any viscous corrections.
\item Boost the particles by $u$
\item Keep or reject the particle with probability $w(p)$. If rejected, continue to the next species.
\end{enumerate}

This procedure should be repeated for each species. However, volume elements tend to be small and there might be many species with small probabilities. Rather than throwing random numbers for each species it is more efficient to implement the following algorithm:
\begin{enumerate}\itemsep=0pt
\item Generate a series of thresholds separated by random amounts, $-\ln({\rm ran}())$, where ${\rm ran}()$ generates a random number between zero and unity. This produces thresholds separated by lengths with probabilities $e^{-x}$. If one is at a position $x$ and one increments by $dx\rightarrow 0$, the chance of hitting a threshold is $dx$, independent of any previous threshold.
\item Thus one can make a cumulative sum for each species $i$,
\begin{equation}
S_i=\sum_{j\le i} \Omega_{\rm opt}n_j,
\end{equation}
where $n_j$ is the density in the rest frame of species $j$. If $S_i$ crosses a threshold, one considers creating a particle as described above. If $S_i$ crosses two thresholds, one attempts to create two particles. 
\end{enumerate}

For each differential probability the chance of crossing a threshold will be independent of any other differential probability, so the emissions are independent, and thus Poissonian. With this approach, the number of random numbers one generates in deciding whether to attempt creating a particle is the number of particles one would make if the weights, $w(p)$ in Eq. (\ref{eq:weightdef}), were unity. This can be a significant improvement compared to throwing random numbers for each species (perhaps over 300) in each volume element (perhaps many millions). Since one knows the total density of hadrons, $n_{\rm had}$, one can check to see whether $n_{\rm had}\Omega_{\rm opt}$ crosses a threshold. If not, one can increment the running total in one swipe and avoid considering each species independently.

This procedure assumes that you already know $n_j$. For calculations where the interfaces is defined by a fixed temperature, and with $\mu=0$, one can calculate these densities once, and store them. For calculations with a variety of breakup densities, one might store an array of values.

\subsection{Generating a particle with the thermal phase space density $f({\bf p})$.}

In the absence of viscous corrections one can generate a particle in the fluid rest frame with the following algorithm. For a probability distribution $\sim x^{n-1}e^{-x}$, one can sample the distribution by taking the natural log of the product of three random numbers, i.e. $x=-\ln(r_1r_2\cdots r_n)$, where $r_i$ are random numbers chosen uniformly between zero and unity. For a three-dimensional distribution of massless particles, the choice of coordinates is
\begin{eqnarray*}
p&=&-T\ln(r_1r_2r_3),\\
\cos\theta&=&\frac{\ln(r_1)-\ln(r_2)}{\ln(r_1)+\ln(r_2)},\\
\phi&=&\frac{2\pi [\ln(r_1r_2)]^2}{[\ln(r_1r_2r_3)]^2}.
\end{eqnarray*}
To check that this works one can calculate  the Jacobian,
\begin{eqnarray}
dN&=&dr_1dr_2dr_3=dpd\cos\theta d\phi|J|,\\
\nonumber
J&=&\frac{p^2}{8\pi T^3}e^{-p/T}.
\end{eqnarray}
For massive particles one can throw a fourth random number $r_4$, and if $r_4>e^{-(E-p)/T}$, one repeats the procedure until one satisfies the additional condition.

Unfortunately, this becomes inefficient for large masses, $m$, because the probability of successfully satisfying the condition for $r_4$ becomes small for small $T/m$. In that case one applies an alternative algorithm. For $T/m<0.6$, it was found that a more efficient method is based on the expression,
\begin{equation}
dN\sim p^2dp~e^{-E_p/T}\sim\frac{p}{E_p}(K+m)^2dK~e^{-K/T},
\end{equation}
where $K=E-m$ is the kinetic energy. The strategy is to generate $K$ ignoring the factor $p/E_p$, then do a keep-or-repeat based on that weight. To generate a value of $K$ consistent with $(K+m)^2e^{-K/t}$, one breaks up the factor $(K+m)^2$ into three terms,
\begin{equation}
dN\sim (K^2+2mK+m^2)e^{-K/T}.
\end{equation}
One first throws a random number $r_0$ and chooses which of the three terms to use as a distribution based on the integrated weights for each term, $2T^3,2mT^2$ and $m^2T$. Once one has picked a given term, one can pick $K$ as $-T\ln(r_1r_2r_3),-T\ln(r_1r_2)$ and $-T\ln(r_1)$ respectively. With this value of $K$, one can now do a keep or repeat decision based on the weight $p/E$. After $K$ is chosen, $\cos\theta$ and $\phi$ can be picked with new random numbers, $r_4$and $r_5$.

Viscous corrections can be applied according to \cite{Pratt:2010jt}. This involves transforming the momentum according to (in the rest frame)
\begin{equation}
p_i=(\delta_{ij}+\gamma \tau_{ij})p_j,
\end{equation}
where $\tau_{ij}$ is the shear tensor in the frame of the fluid, and $\gamma$ is a constant chosen so that the generated distribution will indeed have the stress-energy tensor one wishes assuming that the viscous correction $\tau$ is much smaller than the pressure. This coefficient $\gamma$ can be found analytically given the list of masses and spins of the hadrons \cite{Pratt:2010jt}. For $\tau/P< 1/2$, the linear approximation is good to the one percent level or better in that it consistently reproduces the viscous correction $\tau_{ij}$ according to
\begin{equation}
\tau_{ij}=\frac{1}{\Omega}\sum_i\int{d^3p}~\frac{dN}{d^3p}\frac{p_ip_j}{E_p}.
\end{equation}

For modeling collisions at RHIC or at the LHC, the algorithm described here can generate the thousands of particles needed for a few units of rapidity in a few hundredths of a second.

\begin{acknowledgments}
This work was supported by the Department of Energy Office of Science through grant number DE-FG02-03ER41259.
\end{acknowledgments}

\end{document}